\documentstyle[12pt]{article}
\def \be{\begin{equation}}
\def \ee{\end{equation}}
\def \bea{\begin{eqnarray}}
\def \eea{\end{eqnarray}}

\title{Thermodynamical quantities of lattice full QCD from an efficient method}
\vspace{2cm}
\author{Xiang-Qian Luo\\
{\small\sl Department of Physics, Zhongshan University, 
Guangzhou 510275, China}\\
}

\date{\today}

\begin{document}
\maketitle

\begin{abstract}
I extend to QCD an efficient method for lattice gauge theory with dynamical fermions. 
Once the eigenvalues of the Dirac operator  and 
the density of states of pure gluonic configurations
at a set of plaquette energies (proportional to the gauge action)  are computed,
thermodynamical quantities deriving from the partition function 
can be obtained for arbitrary flavor number,  
quark masses and wide range of coupling constants, without additional computational cost.
Results for the chiral condensate and gauge action
are presented on the $10^4$ lattice at flavor number 
$N_f=0$, 1, 2, 3, 4 and many  quark masses and coupling constants.
New results in the chiral limit for the gauge action and its correlation with the chiral condensate, 
which are useful for
analyzing the QCD chiral phase structure, are also provided. 
\end{abstract}

\section{Introduction}
\label{sec1}


Although it is believed to be the most powerful non-perturbative approach to quantum field theory, 
lattice Monte Carlo techniques still suffer from systematic errors.
The sources of systematic error are mainly finite size effects, finite lattice spacing, 
quenched approximation, and the use of unphysical quark masses.
Finite volume effects and lattice spacing errors are relevant, and can be significantly reduced 
through the implementation of the Symanzik improvement
\cite{Symanzik:1983dc,Hamber:1983qa,Hasenfratz:1994sp,Lepage:1996jw,Luo:1994af,Luo:1996tx,Luo:1999dx,Jiang:1999hk}.
Quenched approximation ignores the feedback of fermions 
(the determinant of the Dirac operator $\Delta$
in the measure is constraint to be $\det \Delta=1$). 
More recent investigations \cite{Sharpe:1992ft,Bernard:1992mk,Aoki:2000yr}
suggest that quenched approximation may lead to a systematic deviation from experiment.

The most popular algorithms for generating unquenched configurations are
the hybrid Monte Carlo (HMC) \cite{Duane:1987de} 
and/or  hybrid molecular dynamics (HMD) algorithms \cite{Gottlieb:1987mq}.
In these algorithms, each trial configuration is generated by solving the equation of motion with
some number of steps, and in each step, numerical inversion of the Dirac matrix has to be done.
For a different set of bare parameters of ``sea" quark mass $m_{sea}$, flavor number $N_f$,
and coupling constant $g$, 
new independent simulations are required. Therefore, these algorithms 
are still very expensive and require high performance supercomputers,
in particular for small quark masses, even on a moderate lattice. 

Without loss of generality, let us discuss the unimproved lattice QCD with Kogut-Susskind quarks,
which has the action
$S=S_g+S_f$.
Here
\begin{eqnarray}
S_g &=& - {\beta \over N_c} \sum_{p} {\rm Re} {\rm Tr} (U_p),
\nonumber \\
S_f &=& \sum_{x,y} {\bar \psi}(x) \Delta_{x,y} \psi (y), 
\nonumber \\
U_p &=& U_{\mu}(x) U_{\nu}(x+\mu) U_{\mu}^{\dagger} (x+\nu) U_{\nu}^{\dagger} (x), 
\nonumber \\
\Delta_{x,y}  &=& m \delta_{x,y}
\nonumber \\
& + &
{1 \over 2} \eta_{\mu}(x)  \left[ U_{\mu}(x) \delta_{x,y-\mu} 
-U_{\mu}^{\dagger} (x-\mu) \delta_{x,y+\mu} \right],
\nonumber \\
\eta_{\mu}(x) &=& (-1)^{x_1+x_2+...+x_{\mu-1}}.
\nonumber \\
\end{eqnarray}
The bare coupling constant $g$ is related to $\beta$ by $\beta=2 N_c/g^2$
with $N_c=3$ the number of colors.
The corresponding partition function is
\begin{eqnarray}
{\cal Z} = \int [dU] ~  {\rm e}^{-S_g} \left(\det \Delta(m, U)\right)^{N_f/4}.
\label{partition}
\end{eqnarray}
In most simulations, 
people (e.g. \cite{Altmeyer:1993dd}) 
usually distinguish ``sea" quarks (for the configurations)  
from the "valence" quarks (for the quark propagators), due to the heavy costs in
generating the full QCD configurations at different set of sea quark masses. 
Let us take the chiral condensate as an example, which is meant to compute  
\begin{eqnarray}
\langle {\bar \psi} \psi \rangle 
&=& {\int [dU] {\rm Tr} \Delta^{-1} (m_{val},U)
e^{-S_g} \left(  \det \Delta (m_{sea},U) \right)^{N_f/4}
\over  VN_c ~ \int [dU] e^{-S_g} \left( \det \Delta (m_{sea},U) \right)^{N_f/4}}
\nonumber \\
&&\label{ps}
\end{eqnarray}
at a set of valence quark mass $m_{val}$, while always
using a full configuration at a given sea quark mass $m_{sea}\not=m_{val}$.
Unfortunately, this distinction is not theoretically
consistent with the physical case $m=m_{val}=m_{sea}$.

In this work, I will describe an efficient method 
for simulating lattice QCD with dynamical fermions,
which is particularly useful for investigating the thermodynamical properties
and chiral properties.
Results will be provided for the chiral condensate and gauge action in QCD 
on the $10^4$ lattice for flavor number $N_f=1, 2, 3, 4$,
and many values of quark mass $m$ and coupling constant $g$. 
New results in the chiral limit 
for the gauge action and its correlation with the chiral condensate
will also be presented.

\section{METHODOLOGY}
\label{sec2}

The method described here is a generalization of the 
microcanonical fermionic average method \cite{Azcioti:1990ng}
to QCD in 3+1 dimensions.
Using the identity
\begin{eqnarray}
{\rm e}^{-S_g} = \int dE ~ \delta \left( {1 \over N_c} \sum_{p} 
{\rm Re} {\rm Tr} (U_p)- 6VE \right) ~ {\rm e}^{6V \beta E},
\end{eqnarray}
one can rewrite the partition function as
a one-dimensional integral
\begin{eqnarray}
{\cal Z} = \int dE ~ e^{-S^{{\rm eff}}(E,m,N_f,\beta)}.
\label{partition1}
\end{eqnarray}
Here $S^{{\rm eff}}$ is the full effective action as a function of plaquette energy $E$,
quark mass $m$, flavor number $N_f$, and coupling $\beta$ defined by 
\begin{eqnarray}
S^{{\rm eff}}(E,m,N_f,\beta) &=& - \ln n(E) 
- 6 \beta VE 
\nonumber \\
&+& S_f^{{\rm eff}}(E,m,N_f),
\label{full}
\end{eqnarray}
where $V$ is the total number of lattice sites,
\begin{eqnarray}
n (E) =  \int [dU] ~
\delta \left( {1 \over N_c} \sum_{p} {\rm Re} {\rm Tr} (U_p)- 6VE \right)
\label{density}
\end{eqnarray}
is the density of states at the given $E$, and
\begin{eqnarray}
S^{{\rm eff}}_f &=& - 
\ln  \langle  \left(\det \Delta(m, U)\right)^{N_f/4}
 \rangle_E
\nonumber \\
&=& -\ln \langle  \left(\prod_{i=1}^{N_cV/2} \left( \lambda_i^2 (U) +m^2\right) \right)^{N_f/4}
 \rangle_E
\label{seff}
\end{eqnarray}
is the effective fermionic action. Non-vanishing $S^{{\rm eff}}_f$ implies the
interaction between quarks and gluons. 
Here $\lambda_i (U)$ is the $i$-th positive eigenvalue 
of the massless Dirac
operator $\Delta(m=0)$ and $\langle ... \rangle_E$ is the average of the observable over configurations
with the probability distribution 
$\delta \left( \sum_{p} {\rm Re} {\rm Tr} (U_p)/N_c- 6VE \right)/n(E)$
at fixed plaquette energy $E$.
$S^{{\rm eff}}_f$ does not depend on $\beta$.
Once I compute the the positive eigenvalues of $\Delta(m=0)$ for {\it pure gauge} configurations 
at a set of $E$, I can obtain at no cost 
$S^{{\rm eff}}_f$ and therefore the partition function for
any $m$, $N_f$ and $\beta$. Then 
the thermodynamical properties 
can be obtained from the derivatives of the partition function.
For example the chiral condensate and the vacuum expectation value of the plaquette are
\begin{eqnarray}
E_p &=& {\langle {\rm Re ~Tr} U_p  \rangle \over N_c}
= {1 \over 6V}  {\partial \ln {\cal Z} \over \partial \beta}
\nonumber \\
&=& {\int dE ~ Ee^{-S^{{\rm eff}}(E,m,N_f,\beta)} \over {\cal Z}},
\nonumber \\
\langle {\bar \psi} \psi \rangle 
&=& {1 \over VN_cN_f/4} {\partial \ln {\cal Z} \over \partial m}
\nonumber \\
&=&{-1 \over VN_cN_f/4}{\int dE ~ {\partial S^{{\rm eff}}_f \over \partial m}
e^{-S^{{\rm eff}}(E,m,N_f,\beta)} \over {\cal Z}}
\nonumber \\
&=& {1 \over VN_c {\cal Z}}
\int d E ~ e^{-S^{{\rm eff}}(E,m,N_f,\beta)}
\nonumber \\
& \times & {
\langle \sum_{i=1}^{N_cV/2} {2m \over \lambda_i^2 (U) +m^2}
\left(\det \Delta(m, U)\right)^{N_f/4}
 \rangle_E
 \over
\langle  \left(\det \Delta(m, U)\right)^{N_f/4}
 \rangle_E
}.
\label{thermo}
\end{eqnarray}
We can also calculate the specific heat, chiral susceptibility and other local quantities.
One prominent feature is that the effective action 
and partition function are also calculable in the chiral
limit, which is very useful for studying the chiral properties.

\section{IMPLEMENTATION}
\label{sec3}

The basic idea of the algorithm described in Sect. \ref{sec2} 
is to compute the effective action in Eq. (\ref{full}). The density of states
$n(E)$ can be directly evaluated by numerically
integrating out the quenched SU(3) data: 
\begin{eqnarray}
- {\ln n(E) \over V} =6 \int_{0}^{E} dE' ~~ \beta (E',N_f=0)+ {\rm const.}.
\label{lnn}
\end{eqnarray}

The most important and time consuming part of the work, is the computation of 
$S^{{\rm eff}}_f$ in Eq. (\ref{seff}) 
by averaging out the fermionic determinant over configurations at the given $E$.
These configurations can be generated by microcanonical or over-relaxation processes.
For SU(3),  although there exists a microcanonical algorithm \cite{Petronzio:1990fb},
it is very difficult to implement it ergodically.
To solve the problem, 
I use a different prescription, 
over-relation-updating the subgroups SU(2) of SU(3)  described as follows. 
Let $U$ be an SU(3) link to be updated, and $R$ the sum of 6 staples.

\noindent
(a)
Find the $2 \times 2$  block of $UR$ by striking out the 3rd row and column
\cite{Cabibbo:1982zn}:
\begin{eqnarray}
  \left( 
    \begin{array}{ccc}
     B_1 & B_3 & \times \\
        B_2 &  B_4 & \times \\
         \times & \times & \times \\
     \end{array}
   \right).
\label{b1}
\end{eqnarray}
Write 
\begin{eqnarray}
B=r_0+i\vec{r} \cdot \vec{\sigma}=
 \left( 
    \begin{array}{cc}
          B_1 & B_3 \\
          B_2 & B_4 \\
     \end{array}
   \right),
\label{b}
\end{eqnarray}
with $r_0$ and $\vec{r}$ being complex.
Then find  $B'$, and the norm $k$:
\begin{eqnarray}
B' &=& {\rm Re}(r_0)+i{\rm Re} (\vec{r}) \cdot \vec{\sigma}
\nonumber \\
&=& 
\left( 
    \begin{array}{cc}
          (B_1+B_4^{\star})/2 & ~~ (B_3-B_2^{\star})/2 \\
          (B_2-B_3^{\star})/2 & ~~  (B_1^{\star}+B_4)/2 \\
     \end{array}
   \right),
\nonumber \\
k &=& \sqrt{\det B'},
\label{b'}
\end{eqnarray}
so that $B'/k$ is a SU(2) matrix.
Denote
\begin{eqnarray}
C=((B')^{\dagger}/k)^2=
 \left( 
    \begin{array}{cc}
          C_1 & C_3 \\
          C_2 & C_4 \\
     \end{array}
   \right), 
\label{C}
\end{eqnarray}
and
\begin{eqnarray}
a_1=  \left( 
    \begin{array}{ccc}
     C_1 & C_3 & 0 \\
     C_2 & C_4  & 0 \\
     0 & 0 & 1 \\
      \end{array}
   \right).
\label{a1}
\end{eqnarray}
Perform over-relaxation updates 
\cite{Brown:1987rr} of the SU(2) subgroups: $U'=a_1 U$.

\noindent
(b)
Find the $2 \times 2$  block of $U'R$ by striking out the 2nd row and column:
\begin{eqnarray}
  \left( 
    \begin{array}{ccc}
     B_1 & \times & B_3\\
         \times & \times & \times \\
        B_2 & \times &  B_4 \\
     \end{array}
   \right),
\label{b2}
\end{eqnarray}
Repeat the procedure of finding $C$ as Eqs. (\ref{b}), (\ref{b'}), (\ref{C}) and denote
\begin{eqnarray}
a_2= 
  \left( 
    \begin{array}{ccc}
      C_1 & 0 & C_3\\
         0 & 1 & 0 \\
        C_2 & 0 &  C_4 \\
     \end{array}
   \right),
\label{a2}
\end{eqnarray}
$U''=a_2 U'$.

\noindent
(c)
Find the $2 \times 2$ block of $U"R$ by striking out the 1st row and column:
\begin{eqnarray}
 \left( 
    \begin{array}{ccc}
     \times & \times & \times\\
         \times & B_1 & B_3 \\
         \times & B_2 & B_4 \\
     \end{array}
   \right).
\label{b3}
\end{eqnarray}
Repeat the procedure of finding $C$ as Eqs. (\ref{b}), (\ref{b'}), (\ref{C}) and denote
\begin{eqnarray}
a_3= 
 \left( 
    \begin{array}{ccc}
     1 & 0 & 0\\
         0 & C_1 & C_3 \\
         0 & C_2 & C_4 \\
     \end{array}
   \right).
\label{a3}
\end{eqnarray}
The new link is now $U_{new}=a_3 U"$. 
The processes above satisfy ${\rm Tr}(U_{new}R)={\rm Tr} (U''R)={\rm Tr}(U'R)={\rm Tr}(UR)$, where
the plaquette energy remains unchanged.

The Lanczos algorithm \cite{Lanczos} was designed  for calculating the eigenvalues of a large sparse
matrix, but the rounding errors grow exponentially for the larger eigenvalues. 
I use the modified Lanczos algorithm \cite{Cullum,Barbour} 
to solve this problem so that all the true eigenvalues can be found.
On a $d$ dimensional lattice with $V$ lattice sites, 
the following sum rules  
\begin{eqnarray}
\sum_{i=1}^{N_cV/2} \lambda_i^2 &=& N_cV,
\nonumber \\
\sum_{i=1}^{N_cV/2} \lambda_i^4 &=& \left[{4d-1 \over 2}- (d-1)E \right] {d N_c^2V \over 8}
\label{sum}
\end{eqnarray}
can be used to check the accuracy of the eigenvalues of $\Delta (m=0)$.

\section{RESULTS}
\label{sec4}

\subsection{Microcanonical updates}

I describe now the QCD data on the $10^4$ lattice, obtained on a workstation. 
For the gauge fields, periodic boundary conditions are used. 
For fermions anti-periodic boundary condition
in the time direction is implemented.

\begin{figure}[htb]
\begin{center}
\setlength{\unitlength}{0.240900pt}
\ifx\plotpoint\undefined\newsavebox{\plotpoint}\fi
\sbox{\plotpoint}{\rule[-0.200pt]{0.400pt}{0.400pt}}%
\begin{picture}(1500,900)(0,0)
\font\gnuplot=cmr10 at 10pt
\gnuplot
\sbox{\plotpoint}{\rule[-0.200pt]{0.400pt}{0.400pt}}%
\put(161.0,123.0){\rule[-0.200pt]{4.818pt}{0.400pt}}
\put(141,123){\makebox(0,0)[r]{0}}
\put(1419.0,123.0){\rule[-0.200pt]{4.818pt}{0.400pt}}
\put(161.0,254.0){\rule[-0.200pt]{4.818pt}{0.400pt}}
\put(141,254){\makebox(0,0)[r]{0.2}}
\put(1419.0,254.0){\rule[-0.200pt]{4.818pt}{0.400pt}}
\put(161.0,385.0){\rule[-0.200pt]{4.818pt}{0.400pt}}
\put(141,385){\makebox(0,0)[r]{0.4}}
\put(1419.0,385.0){\rule[-0.200pt]{4.818pt}{0.400pt}}
\put(161.0,515.0){\rule[-0.200pt]{4.818pt}{0.400pt}}
\put(141,515){\makebox(0,0)[r]{0.6}}
\put(1419.0,515.0){\rule[-0.200pt]{4.818pt}{0.400pt}}
\put(161.0,646.0){\rule[-0.200pt]{4.818pt}{0.400pt}}
\put(141,646){\makebox(0,0)[r]{0.8}}
\put(1419.0,646.0){\rule[-0.200pt]{4.818pt}{0.400pt}}
\put(161.0,777.0){\rule[-0.200pt]{4.818pt}{0.400pt}}
\put(141,777){\makebox(0,0)[r]{1}}
\put(1419.0,777.0){\rule[-0.200pt]{4.818pt}{0.400pt}}
\put(161.0,123.0){\rule[-0.200pt]{0.400pt}{4.818pt}}
\put(161,82){\makebox(0,0){0}}
\put(161.0,757.0){\rule[-0.200pt]{0.400pt}{4.818pt}}
\put(321.0,123.0){\rule[-0.200pt]{0.400pt}{4.818pt}}
\put(321,82){\makebox(0,0){5}}
\put(321.0,757.0){\rule[-0.200pt]{0.400pt}{4.818pt}}
\put(480.0,123.0){\rule[-0.200pt]{0.400pt}{4.818pt}}
\put(480,82){\makebox(0,0){10}}
\put(480.0,757.0){\rule[-0.200pt]{0.400pt}{4.818pt}}
\put(640.0,123.0){\rule[-0.200pt]{0.400pt}{4.818pt}}
\put(640,82){\makebox(0,0){15}}
\put(640.0,757.0){\rule[-0.200pt]{0.400pt}{4.818pt}}
\put(800.0,123.0){\rule[-0.200pt]{0.400pt}{4.818pt}}
\put(800,82){\makebox(0,0){20}}
\put(800.0,757.0){\rule[-0.200pt]{0.400pt}{4.818pt}}
\put(960.0,123.0){\rule[-0.200pt]{0.400pt}{4.818pt}}
\put(960,82){\makebox(0,0){25}}
\put(960.0,757.0){\rule[-0.200pt]{0.400pt}{4.818pt}}
\put(1119.0,123.0){\rule[-0.200pt]{0.400pt}{4.818pt}}
\put(1119,82){\makebox(0,0){30}}
\put(1119.0,757.0){\rule[-0.200pt]{0.400pt}{4.818pt}}
\put(1279.0,123.0){\rule[-0.200pt]{0.400pt}{4.818pt}}
\put(1279,82){\makebox(0,0){35}}
\put(1279.0,757.0){\rule[-0.200pt]{0.400pt}{4.818pt}}
\put(1439.0,123.0){\rule[-0.200pt]{0.400pt}{4.818pt}}
\put(1439,82){\makebox(0,0){40}}
\put(1439.0,757.0){\rule[-0.200pt]{0.400pt}{4.818pt}}
\put(161.0,123.0){\rule[-0.200pt]{307.870pt}{0.400pt}}
\put(1439.0,123.0){\rule[-0.200pt]{0.400pt}{157.549pt}}
\put(161.0,777.0){\rule[-0.200pt]{307.870pt}{0.400pt}}
\put(40,450){\makebox(0,0){$E_p$}}
\put(800,21){\makebox(0,0){$\beta$}}
\put(800,839){\makebox(0,0){$N_f=0$}}
\put(161.0,123.0){\rule[-0.200pt]{0.400pt}{157.549pt}}
\put(167,130){\raisebox{-.8pt}{\makebox(0,0){$\Diamond$}}}
\put(180,146){\raisebox{-.8pt}{\makebox(0,0){$\Diamond$}}}
\put(193,162){\raisebox{-.8pt}{\makebox(0,0){$\Diamond$}}}
\put(206,180){\raisebox{-.8pt}{\makebox(0,0){$\Diamond$}}}
\put(219,198){\raisebox{-.8pt}{\makebox(0,0){$\Diamond$}}}
\put(231,217){\raisebox{-.8pt}{\makebox(0,0){$\Diamond$}}}
\put(244,237){\raisebox{-.8pt}{\makebox(0,0){$\Diamond$}}}
\put(257,257){\raisebox{-.8pt}{\makebox(0,0){$\Diamond$}}}
\put(270,278){\raisebox{-.8pt}{\makebox(0,0){$\Diamond$}}}
\put(282,301){\raisebox{-.8pt}{\makebox(0,0){$\Diamond$}}}
\put(295,325){\raisebox{-.8pt}{\makebox(0,0){$\Diamond$}}}
\put(308,352){\raisebox{-.8pt}{\makebox(0,0){$\Diamond$}}}
\put(321,385){\raisebox{-.8pt}{\makebox(0,0){$\Diamond$}}}
\put(334,431){\raisebox{-.8pt}{\makebox(0,0){$\Diamond$}}}
\put(346,494){\raisebox{-.8pt}{\makebox(0,0){$\Diamond$}}}
\put(359,524){\raisebox{-.8pt}{\makebox(0,0){$\Diamond$}}}
\put(372,545){\raisebox{-.8pt}{\makebox(0,0){$\Diamond$}}}
\put(385,562){\raisebox{-.8pt}{\makebox(0,0){$\Diamond$}}}
\put(397,576){\raisebox{-.8pt}{\makebox(0,0){$\Diamond$}}}
\put(410,589){\raisebox{-.8pt}{\makebox(0,0){$\Diamond$}}}
\put(423,600){\raisebox{-.8pt}{\makebox(0,0){$\Diamond$}}}
\put(436,609){\raisebox{-.8pt}{\makebox(0,0){$\Diamond$}}}
\put(449,618){\raisebox{-.8pt}{\makebox(0,0){$\Diamond$}}}
\put(461,625){\raisebox{-.8pt}{\makebox(0,0){$\Diamond$}}}
\put(474,632){\raisebox{-.8pt}{\makebox(0,0){$\Diamond$}}}
\put(480,635){\raisebox{-.8pt}{\makebox(0,0){$\Diamond$}}}
\put(512,649){\raisebox{-.8pt}{\makebox(0,0){$\Diamond$}}}
\put(544,661){\raisebox{-.8pt}{\makebox(0,0){$\Diamond$}}}
\put(576,670){\raisebox{-.8pt}{\makebox(0,0){$\Diamond$}}}
\put(608,679){\raisebox{-.8pt}{\makebox(0,0){$\Diamond$}}}
\put(640,685){\raisebox{-.8pt}{\makebox(0,0){$\Diamond$}}}
\put(672,692){\raisebox{-.8pt}{\makebox(0,0){$\Diamond$}}}
\put(704,697){\raisebox{-.8pt}{\makebox(0,0){$\Diamond$}}}
\put(736,701){\raisebox{-.8pt}{\makebox(0,0){$\Diamond$}}}
\put(768,706){\raisebox{-.8pt}{\makebox(0,0){$\Diamond$}}}
\put(800,709){\raisebox{-.8pt}{\makebox(0,0){$\Diamond$}}}
\put(832,713){\raisebox{-.8pt}{\makebox(0,0){$\Diamond$}}}
\put(896,718){\raisebox{-.8pt}{\makebox(0,0){$\Diamond$}}}
\put(960,723){\raisebox{-.8pt}{\makebox(0,0){$\Diamond$}}}
\put(1024,727){\raisebox{-.8pt}{\makebox(0,0){$\Diamond$}}}
\put(1088,731){\raisebox{-.8pt}{\makebox(0,0){$\Diamond$}}}
\put(1151,734){\raisebox{-.8pt}{\makebox(0,0){$\Diamond$}}}
\put(1215,737){\raisebox{-.8pt}{\makebox(0,0){$\Diamond$}}}
\put(1279,739){\raisebox{-.8pt}{\makebox(0,0){$\Diamond$}}}
\put(1343,741){\raisebox{-.8pt}{\makebox(0,0){$\Diamond$}}}
\put(1407,743){\raisebox{-.8pt}{\makebox(0,0){$\Diamond$}}}
\end{picture}
\end{center}
\caption{The vacuum expectation value of the plaquette energy from quenched simulation.}
\label{fig1}
\end{figure}
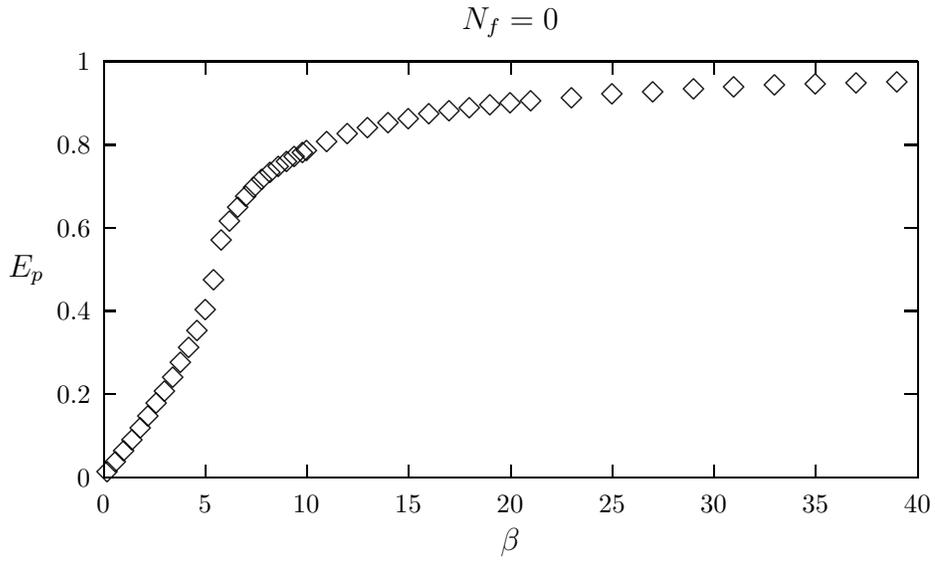

First,  I calculate the quenched mean plaquette energy as a function of $\beta$, 
using the Cabibbo-Marinari \cite{Cabibbo:1982zn} algorithm.
High precision for this quantity can be easily achieved. 
Figure \ref{fig1} shows the result.
Then I calculate the density of states using Eq. (\ref{lnn}), by numerically
integrating out the interpolated data for $\beta (E)$ according to the trapezoidal rule. 
The result is shown in Fig. \ref{fig2}, while the irrelevant constant in Eq. (\ref{lnn})
is ignored.

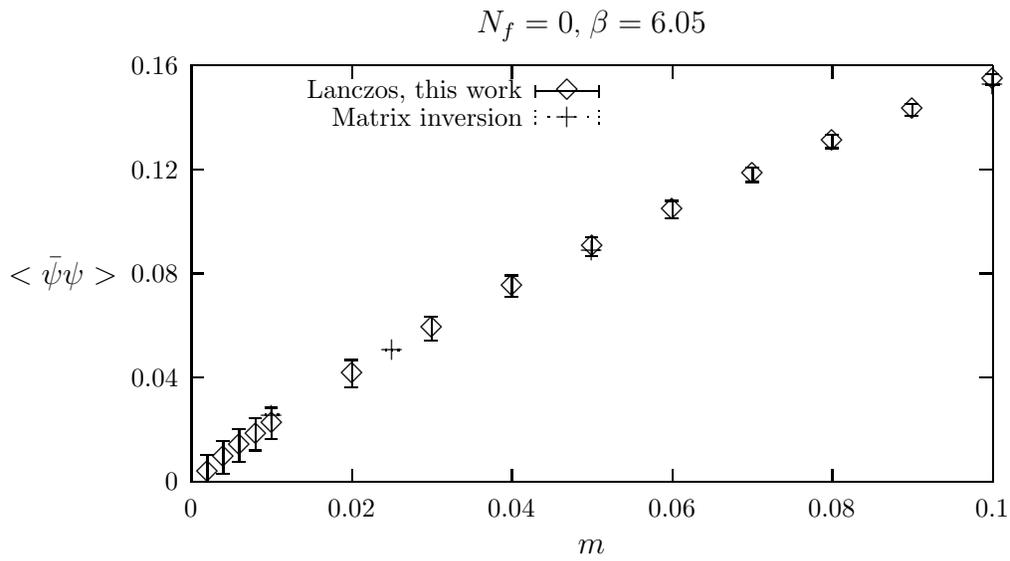
\begin{figure}
\begin{center}
\setlength{\unitlength}{0.240900pt}
\ifx\plotpoint\undefined\newsavebox{\plotpoint}\fi
\sbox{\plotpoint}{\rule[-0.200pt]{0.400pt}{0.400pt}}%
\begin{picture}(1500,900)(0,0)
\font\gnuplot=cmr10 at 10pt
\gnuplot
\sbox{\plotpoint}{\rule[-0.200pt]{0.400pt}{0.400pt}}%
\put(181.0,123.0){\rule[-0.200pt]{4.818pt}{0.400pt}}
\put(161,123){\makebox(0,0)[r]{0}}
\put(1419.0,123.0){\rule[-0.200pt]{4.818pt}{0.400pt}}
\put(181.0,287.0){\rule[-0.200pt]{4.818pt}{0.400pt}}
\put(161,287){\makebox(0,0)[r]{0.04}}
\put(1419.0,287.0){\rule[-0.200pt]{4.818pt}{0.400pt}}
\put(181.0,450.0){\rule[-0.200pt]{4.818pt}{0.400pt}}
\put(161,450){\makebox(0,0)[r]{0.08}}
\put(1419.0,450.0){\rule[-0.200pt]{4.818pt}{0.400pt}}
\put(181.0,613.0){\rule[-0.200pt]{4.818pt}{0.400pt}}
\put(161,613){\makebox(0,0)[r]{0.12}}
\put(1419.0,613.0){\rule[-0.200pt]{4.818pt}{0.400pt}}
\put(181.0,777.0){\rule[-0.200pt]{4.818pt}{0.400pt}}
\put(161,777){\makebox(0,0)[r]{0.16}}
\put(1419.0,777.0){\rule[-0.200pt]{4.818pt}{0.400pt}}
\put(181.0,123.0){\rule[-0.200pt]{0.400pt}{4.818pt}}
\put(181,82){\makebox(0,0){0}}
\put(181.0,757.0){\rule[-0.200pt]{0.400pt}{4.818pt}}
\put(433.0,123.0){\rule[-0.200pt]{0.400pt}{4.818pt}}
\put(433,82){\makebox(0,0){0.02}}
\put(433.0,757.0){\rule[-0.200pt]{0.400pt}{4.818pt}}
\put(684.0,123.0){\rule[-0.200pt]{0.400pt}{4.818pt}}
\put(684,82){\makebox(0,0){0.04}}
\put(684.0,757.0){\rule[-0.200pt]{0.400pt}{4.818pt}}
\put(936.0,123.0){\rule[-0.200pt]{0.400pt}{4.818pt}}
\put(936,82){\makebox(0,0){0.06}}
\put(936.0,757.0){\rule[-0.200pt]{0.400pt}{4.818pt}}
\put(1187.0,123.0){\rule[-0.200pt]{0.400pt}{4.818pt}}
\put(1187,82){\makebox(0,0){0.08}}
\put(1187.0,757.0){\rule[-0.200pt]{0.400pt}{4.818pt}}
\put(1439.0,123.0){\rule[-0.200pt]{0.400pt}{4.818pt}}
\put(1439,82){\makebox(0,0){0.1}}
\put(1439.0,757.0){\rule[-0.200pt]{0.400pt}{4.818pt}}
\put(181.0,123.0){\rule[-0.200pt]{303.052pt}{0.400pt}}
\put(1439.0,123.0){\rule[-0.200pt]{0.400pt}{157.549pt}}
\put(181.0,777.0){\rule[-0.200pt]{303.052pt}{0.400pt}}
\put(-20,450){\makebox(0,0){$<\bar{\psi} \psi>$}}
\put(810,21){\makebox(0,0){$m$}}
\put(810,839){\makebox(0,0){$N_f=0$, $\beta=6.05$}}
\put(181.0,123.0){\rule[-0.200pt]{0.400pt}{157.549pt}}
\put(701,737){\makebox(0,0)[r]{Lanczos, this work}}
\put(721.0,737.0){\rule[-0.200pt]{24.090pt}{0.400pt}}
\put(721.0,727.0){\rule[-0.200pt]{0.400pt}{4.818pt}}
\put(821.0,727.0){\rule[-0.200pt]{0.400pt}{4.818pt}}
\put(206.0,123.0){\rule[-0.200pt]{0.400pt}{10.118pt}}
\put(196.0,123.0){\rule[-0.200pt]{4.818pt}{0.400pt}}
\put(196.0,165.0){\rule[-0.200pt]{4.818pt}{0.400pt}}
\put(231.0,135.0){\rule[-0.200pt]{0.400pt}{12.527pt}}
\put(221.0,135.0){\rule[-0.200pt]{4.818pt}{0.400pt}}
\put(221.0,187.0){\rule[-0.200pt]{4.818pt}{0.400pt}}
\put(256.0,154.0){\rule[-0.200pt]{0.400pt}{12.527pt}}
\put(246.0,154.0){\rule[-0.200pt]{4.818pt}{0.400pt}}
\put(246.0,206.0){\rule[-0.200pt]{4.818pt}{0.400pt}}
\put(282.0,172.0){\rule[-0.200pt]{0.400pt}{12.286pt}}
\put(272.0,172.0){\rule[-0.200pt]{4.818pt}{0.400pt}}
\put(272.0,223.0){\rule[-0.200pt]{4.818pt}{0.400pt}}
\put(307.0,190.0){\rule[-0.200pt]{0.400pt}{11.804pt}}
\put(297.0,190.0){\rule[-0.200pt]{4.818pt}{0.400pt}}
\put(297.0,239.0){\rule[-0.200pt]{4.818pt}{0.400pt}}
\put(433.0,271.0){\rule[-0.200pt]{0.400pt}{10.359pt}}
\put(423.0,271.0){\rule[-0.200pt]{4.818pt}{0.400pt}}
\put(423.0,314.0){\rule[-0.200pt]{4.818pt}{0.400pt}}
\put(558.0,345.0){\rule[-0.200pt]{0.400pt}{8.913pt}}
\put(548.0,345.0){\rule[-0.200pt]{4.818pt}{0.400pt}}
\put(548.0,382.0){\rule[-0.200pt]{4.818pt}{0.400pt}}
\put(684.0,413.0){\rule[-0.200pt]{0.400pt}{8.191pt}}
\put(674.0,413.0){\rule[-0.200pt]{4.818pt}{0.400pt}}
\put(674.0,447.0){\rule[-0.200pt]{4.818pt}{0.400pt}}
\put(810.0,477.0){\rule[-0.200pt]{0.400pt}{7.227pt}}
\put(800.0,477.0){\rule[-0.200pt]{4.818pt}{0.400pt}}
\put(800.0,507.0){\rule[-0.200pt]{4.818pt}{0.400pt}}
\put(936.0,537.0){\rule[-0.200pt]{0.400pt}{6.504pt}}
\put(926.0,537.0){\rule[-0.200pt]{4.818pt}{0.400pt}}
\put(926.0,564.0){\rule[-0.200pt]{4.818pt}{0.400pt}}
\put(1062.0,594.0){\rule[-0.200pt]{0.400pt}{5.541pt}}
\put(1052.0,594.0){\rule[-0.200pt]{4.818pt}{0.400pt}}
\put(1052.0,617.0){\rule[-0.200pt]{4.818pt}{0.400pt}}
\put(1187.0,647.0){\rule[-0.200pt]{0.400pt}{5.059pt}}
\put(1177.0,647.0){\rule[-0.200pt]{4.818pt}{0.400pt}}
\put(1177.0,668.0){\rule[-0.200pt]{4.818pt}{0.400pt}}
\put(1313.0,698.0){\rule[-0.200pt]{0.400pt}{4.577pt}}
\put(1303.0,698.0){\rule[-0.200pt]{4.818pt}{0.400pt}}
\put(1303.0,717.0){\rule[-0.200pt]{4.818pt}{0.400pt}}
\put(1439.0,746.0){\rule[-0.200pt]{0.400pt}{4.095pt}}
\put(1429.0,746.0){\rule[-0.200pt]{4.818pt}{0.400pt}}
\put(206,137){\raisebox{-.8pt}{\makebox(0,0){$\Diamond$}}}
\put(231,161){\raisebox{-.8pt}{\makebox(0,0){$\Diamond$}}}
\put(256,180){\raisebox{-.8pt}{\makebox(0,0){$\Diamond$}}}
\put(282,197){\raisebox{-.8pt}{\makebox(0,0){$\Diamond$}}}
\put(307,214){\raisebox{-.8pt}{\makebox(0,0){$\Diamond$}}}
\put(433,292){\raisebox{-.8pt}{\makebox(0,0){$\Diamond$}}}
\put(558,364){\raisebox{-.8pt}{\makebox(0,0){$\Diamond$}}}
\put(684,430){\raisebox{-.8pt}{\makebox(0,0){$\Diamond$}}}
\put(810,492){\raisebox{-.8pt}{\makebox(0,0){$\Diamond$}}}
\put(936,550){\raisebox{-.8pt}{\makebox(0,0){$\Diamond$}}}
\put(1062,606){\raisebox{-.8pt}{\makebox(0,0){$\Diamond$}}}
\put(1187,658){\raisebox{-.8pt}{\makebox(0,0){$\Diamond$}}}
\put(1313,707){\raisebox{-.8pt}{\makebox(0,0){$\Diamond$}}}
\put(1439,755){\raisebox{-.8pt}{\makebox(0,0){$\Diamond$}}}
\put(771,737){\raisebox{-.8pt}{\makebox(0,0){$\Diamond$}}}
\put(1429.0,763.0){\rule[-0.200pt]{4.818pt}{0.400pt}}
\put(701,696){\makebox(0,0)[r]{Matrix inversion}}
\multiput(721,696)(20.756,0.000){5}{\usebox{\plotpoint}}
\put(821,696){\usebox{\plotpoint}}
\put(721.00,706.00){\usebox{\plotpoint}}
\put(721,686){\usebox{\plotpoint}}
\put(821.00,706.00){\usebox{\plotpoint}}
\put(821,686){\usebox{\plotpoint}}
\put(307.00,228.00){\usebox{\plotpoint}}
\put(307,229){\usebox{\plotpoint}}
\put(297.00,228.00){\usebox{\plotpoint}}
\put(317,228){\usebox{\plotpoint}}
\put(297.00,229.00){\usebox{\plotpoint}}
\put(317,229){\usebox{\plotpoint}}
\put(496.00,330.00){\usebox{\plotpoint}}
\put(496,331){\usebox{\plotpoint}}
\put(486.00,330.00){\usebox{\plotpoint}}
\put(506,330){\usebox{\plotpoint}}
\put(486.00,331.00){\usebox{\plotpoint}}
\put(506,331){\usebox{\plotpoint}}
\put(810,487){\usebox{\plotpoint}}
\put(810,487){\usebox{\plotpoint}}
\put(800.00,487.00){\usebox{\plotpoint}}
\put(820,487){\usebox{\plotpoint}}
\put(800.00,487.00){\usebox{\plotpoint}}
\put(820,487){\usebox{\plotpoint}}
\put(1439.00,746.00){\usebox{\plotpoint}}
\put(1439,747){\usebox{\plotpoint}}
\put(1429.00,746.00){\usebox{\plotpoint}}
\put(1449,746){\usebox{\plotpoint}}
\put(1429.00,747.00){\usebox{\plotpoint}}
\put(1449,747){\usebox{\plotpoint}}
\put(307,228){\makebox(0,0){$+$}}
\put(496,331){\makebox(0,0){$+$}}
\put(810,487){\makebox(0,0){$+$}}
\put(1439,747){\makebox(0,0){$+$}}
\put(771,696){\makebox(0,0){$+$}}
\end{picture}
\end{center}
\caption{The quenched chiral condensate $<{\bar \psi} \psi>$ versus $m$ at $\beta=6.05$.}
\label{fig3}
\end{figure}

At each fixed $E$, 10000-40000 pure gauge configurations are generated, each one is separated by
100 over-relation updates. 100-400 de-correlated configurations are used for the diagonalization of 
the massless Kogut-Susskind fermionic matrix  $\Delta (m=0)$. Their eigenvalues are stored on a disk
with double precision, and the relative errors 
of the sum rules in Eq. (\ref{sum})
are the order of $10^{-8}$ and $10^{-7}$ respectively.
Figure \ref{fig3} plots the quenched chiral condensate 
\begin{eqnarray}
\langle {\bar \psi} \psi \rangle 
&=& {\int dE ~ n(E) ~ e^{6 \beta VE} ~
\langle \sum_{i=1}^{N_cV/2} {2m \over \lambda_i^2 (U) +m^2}
 \rangle_E  \over  VN_c ~ \int dE ~ n(E) ~ e^{6 \beta VE}
}
\end{eqnarray}
versus $m$ at $\beta=6.05$. 
The results are in good agreement with Chen's data \cite{Chen},
although Chen used the fermionic matrix inversion method
on a much larger lattice $16^3 \times 32$.
This provides an additional check of the eigenvalues.

From the eigenvalues of $\Delta (m=0)$ at 16 values of $E \in [0,1)$, 
we can compute the effective fermionic action in Eq. (\ref{seff})
at any $m$ and $N_f$ at almost no cost.
Figures \ref{fig4},\ref{fig5},\ref{fig6}, and \ref{fig7} 
show the effective fermionic action (normalized by the volume)
versus $E$ for $N_f=1, 2, 3$ and 4 respectively. 
I calculated the effective fermionic action for 15 bare quark masses in $m \in [0,0.1]$,
but only illustrate 3 of them.
Statistical errors are less than $O(1/V)=O(10^{-5})$ and invisible at the scale of the figures.
The shapes of the curves are quite similar, but the scales are quite different. 
The slope is  small for small $E<0.4$ and large $E>0.7$, but it is big for intermediate $E \in (0.4,0.7)$. 
We find $S_f^{{\rm eff}} \propto N_f$, which means the effects of the dynamical quarks are proportional to 
the flavor number, and are significant for intermediate $E$.

With the density of states and the effective fermionic action, we can construct
the full effective action in Eq. (\ref{full}) as a function of $E$ for any given $N_f$, $m$ and $\beta$,
using the Newton polynomial interpolation.

\subsection{Themodynamical observables with sea quarks}

The thermodynamical quantities can be obtained 
by numerically integrating out the one-dimensional integrals in Eq. (\ref{thermo}).
 
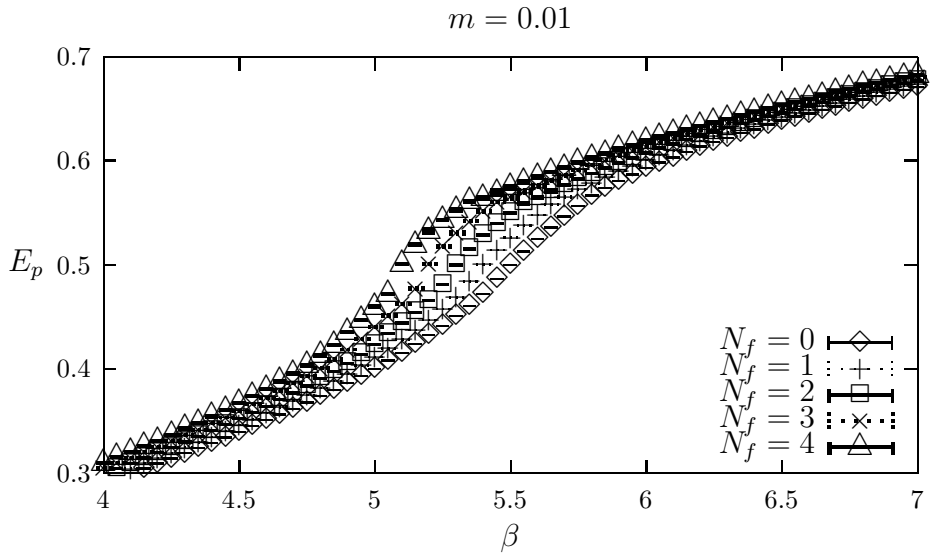
The mean plaquette energy $E_p$ versus
$\beta$ is plotted in 
Fig. \ref{fig8} for $m=m_{sea}=0.01$ and  different $N_f$.
For $N_f=2$ and $\beta=5.7$, $E_p=0.5771624 \pm 6.2251091 \times 10^{-4}$, consistent with 
Chen's HMD data $E_p=0.577386(17)$ on the $16^3 \times 40$ in Ref. \cite{Chen}.
For $N_f=4$ and $\beta=5.4$, $E_p=0.5642520 \pm 8.0579519 \times 10^{-4}$, consistent with 
Chen's HMC data $E_p=0.560334(15)$ on the $16^3 \times 32$ in Ref. \cite{Chen}.
This also means that finite size effects on $E_p$ are not significant.
As seen in this figure, at some fixed $\beta$,
$E_p$ increases with $N_f$. 
The sea quark effects become most
important around $E_p \approx 0.5$.
We can understand this phenomenon by looking at Figs. \ref{fig4},\ref{fig5},\ref{fig6}
and \ref{fig8}
from which one observes maximum slope around $E \approx 0.5$.
The lighter the quark mass,
the bigger the slope is.
Therefore, we have a mechanism: the sea quark effects  
reach their maximum where the effective fermionic action
versus the gauge energy has a maximum slope. 

Figure \ref{fig9} shows the data for $E_p$ in the chiral limit $m=m_{sea}=0$ and different $N_f$.
This quantity has been very useful for analyzing the chiral properties 
of the QED system \cite{Azcoiti:1992}. We believe it will also be useful for QCD.

Figure \ref{fig10} depicts $<{\bar \psi} \psi>$ versus $m$ at $\beta=5.5$
for different $N_f$. The data from the Langevin algorithm   
\cite{Fukugita:1987qb} on the $8^3 \times 18$ lattice
are also included for comparison. (The minor difference between our  $N_f=2$ data  and 
those in \cite{Fukugita:1987qb} might be accounted by 
different lattice sizes used  or systematic error involved). 
Again, one sees $<{\bar \psi} \psi>$
decreases significantly with $N_f$.

\subsection{Correlation between the chiral condensate and gauge action in the chiral limit}

The correlation function  between the plaquette and the chiral condensate  indicates the interaction
between sea quarks and gluons. In the quenched case, it is zero, while for full QCD, it is not.
It can be computed by measuring the mass derivative of the plaquette mean value, since
\begin{eqnarray}
{\partial E_p \over \partial m} 
=
{VN_cN_f \over 4}
\left({\langle \bar{\psi} \psi ~ {\rm Re ~Tr} U_p   \rangle \over N_c} 
- \langle \bar{\psi} \psi  \rangle E_p
\right)
\label{correlation}
\end{eqnarray}
where
\begin{eqnarray}
\bar{\psi} \psi  ={1 \over VN_c} {\rm Tr} \Delta^{-1}.
\label{psi}
\end{eqnarray}
A direct  calculation of the r.h.s. of Eq. (\ref{correlation}) needs very high statistics. 
Also it is impossible to calculate the r.h.s. when $m=0$
because on a finite lattice 
$\langle \bar{\psi} \psi \rangle$ vanishes identically in this limit.
However, 
the l.h.s. of Eq. (\ref{correlation})  can easily be calculated 
in our method. Figure \ref{fig11} plots the results $E_p$ versus $m$
for different $\beta$ and $N_f$.
The mass derivative is obtained by a linear fit to the data, which results
at $m=0$ are given in Tab. I.

\section{OUTLOOK}
\label{sec5}

In this paper, I have extended the microcanonical fermionic average method \cite{Azcioti:1990ng}
 to QCD in 3+1 dimensions,
which allows us to search the parameter space $(N_f,\beta,m)$ with much lower computational cost.
I have provided new data for the mean value of chiral condensates $<{\bar \psi} \psi>$,  
plaquette energy $E_p$, and their correlation function, including the results in the chiral limit
for the latter two quantities.

The disadvantages of the method is that
it is not easy to calculate physical observables 
beyond the thermodynamical quantities (e.g. the spectrum).
The storage of the eigenvalues of the fermionic matrix needs big hard disk space.
Although the method is applicable to the chiral limit, the systematic errors become larger when 
the flavor number and $\beta$ are larger, and the quark mass is smaller.

As the algorithm  \cite{Azcioti:1990ng} has been useful for analyzing the QED phase structure,
I believe the method developed in this work will also be useful for QCD.
Since the algorithm also works in the chiral limit,
the study of spontaneous chiral symmetry breaking \cite{Azcoiti:1995dq}
will be an interesting application of this work
and will be reported elsewhere.

\vskip 1cm

\noindent
{\bf Acknowledgments}

I thank D. Chen for his HMC and HMD data, 
which have been used for
comparison with the work done in this paper. 
I am grateful to E.B. Gregory for useful discussions.
This work is supported by the projects of
National Science Fund for Distinguished Young Scholars (19825117),
National Natural Science Foundation,
Guangdong Provincial Natural Science Foundation (990212), Ministry of Education,
Ministry of Science and Technology,
and the Foundation of Zhongshan University Advanced Research Center.

\begin{table}[hbt]
\caption{Mass derivative of the mean plaquette energy at $m=0$ for $\beta=5.2$.}
\begin{center}
\begin{tabular}{|c|c|}
\hline
$N_f$ & $\partial E_p / \partial m$  \\
\hline
0   &  0\\
1     &  $-0.0357841  \pm 0.0003783$\\
2     &  $-0.0815237   \pm 0.001083$\\
3     &  $-0.327659       \pm  0.0182$\\
4     & $-0.306373        \pm 0.000517$\\
\hline
\end{tabular}
\end{center}
\end{table}

\newpage 

\input fig2.tex
\begin{figure}[htb]
\begin{center}
\setlength{\unitlength}{0.240900pt}
\ifx\plotpoint\undefined\newsavebox{\plotpoint}\fi
\sbox{\plotpoint}{\rule[-0.200pt]{0.400pt}{0.400pt}}%
\begin{picture}(1500,900)(0,0)
\font\gnuplot=cmr10 at 10pt
\gnuplot
\sbox{\plotpoint}{\rule[-0.200pt]{0.400pt}{0.400pt}}%
\put(181.0,123.0){\rule[-0.200pt]{4.818pt}{0.400pt}}
\put(161,123){\makebox(0,0)[r]{-0.1}}
\put(1419.0,123.0){\rule[-0.200pt]{4.818pt}{0.400pt}}
\put(181.0,310.0){\rule[-0.200pt]{4.818pt}{0.400pt}}
\put(161,310){\makebox(0,0)[r]{0}}
\put(1419.0,310.0){\rule[-0.200pt]{4.818pt}{0.400pt}}
\put(181.0,497.0){\rule[-0.200pt]{4.818pt}{0.400pt}}
\put(161,497){\makebox(0,0)[r]{0.1}}
\put(1419.0,497.0){\rule[-0.200pt]{4.818pt}{0.400pt}}
\put(181.0,684.0){\rule[-0.200pt]{4.818pt}{0.400pt}}
\put(161,684){\makebox(0,0)[r]{0.2}}
\put(1419.0,684.0){\rule[-0.200pt]{4.818pt}{0.400pt}}
\put(181.0,123.0){\rule[-0.200pt]{0.400pt}{4.818pt}}
\put(181,82){\makebox(0,0){0}}
\put(181.0,757.0){\rule[-0.200pt]{0.400pt}{4.818pt}}
\put(433.0,123.0){\rule[-0.200pt]{0.400pt}{4.818pt}}
\put(433,82){\makebox(0,0){0.2}}
\put(433.0,757.0){\rule[-0.200pt]{0.400pt}{4.818pt}}
\put(684.0,123.0){\rule[-0.200pt]{0.400pt}{4.818pt}}
\put(684,82){\makebox(0,0){0.4}}
\put(684.0,757.0){\rule[-0.200pt]{0.400pt}{4.818pt}}
\put(936.0,123.0){\rule[-0.200pt]{0.400pt}{4.818pt}}
\put(936,82){\makebox(0,0){0.6}}
\put(936.0,757.0){\rule[-0.200pt]{0.400pt}{4.818pt}}
\put(1187.0,123.0){\rule[-0.200pt]{0.400pt}{4.818pt}}
\put(1187,82){\makebox(0,0){0.8}}
\put(1187.0,757.0){\rule[-0.200pt]{0.400pt}{4.818pt}}
\put(1439.0,123.0){\rule[-0.200pt]{0.400pt}{4.818pt}}
\put(1439,82){\makebox(0,0){1}}
\put(1439.0,757.0){\rule[-0.200pt]{0.400pt}{4.818pt}}
\put(181.0,123.0){\rule[-0.200pt]{303.052pt}{0.400pt}}
\put(1439.0,123.0){\rule[-0.200pt]{0.400pt}{157.549pt}}
\put(181.0,777.0){\rule[-0.200pt]{303.052pt}{0.400pt}}
\put(40,450){\makebox(0,0){$-{S^{\rm m{eff}}_f \over V}$}}
\put(810,21){\makebox(0,0){$E$}}
\put(810,839){\makebox(0,0){$N_f=1$}}
\put(181.0,123.0){\rule[-0.200pt]{0.400pt}{157.549pt}}
\put(361,737){\makebox(0,0)[r]{$m=0.10$}}
\put(185,233){\raisebox{-.8pt}{\makebox(0,0){$\Diamond$}}}
\put(288,259){\raisebox{-.8pt}{\makebox(0,0){$\Diamond$}}}
\put(394,289){\raisebox{-.8pt}{\makebox(0,0){$\Diamond$}}}
\put(497,323){\raisebox{-.8pt}{\makebox(0,0){$\Diamond$}}}
\put(601,362){\raisebox{-.8pt}{\makebox(0,0){$\Diamond$}}}
\put(642,380){\raisebox{-.8pt}{\makebox(0,0){$\Diamond$}}}
\put(680,398){\raisebox{-.8pt}{\makebox(0,0){$\Diamond$}}}
\put(723,420){\raisebox{-.8pt}{\makebox(0,0){$\Diamond$}}}
\put(765,445){\raisebox{-.8pt}{\makebox(0,0){$\Diamond$}}}
\put(807,472){\raisebox{-.8pt}{\makebox(0,0){$\Diamond$}}}
\put(849,502){\raisebox{-.8pt}{\makebox(0,0){$\Diamond$}}}
\put(890,532){\raisebox{-.8pt}{\makebox(0,0){$\Diamond$}}}
\put(1016,602){\raisebox{-.8pt}{\makebox(0,0){$\Diamond$}}}
\put(1120,645){\raisebox{-.8pt}{\makebox(0,0){$\Diamond$}}}
\put(1230,684){\raisebox{-.8pt}{\makebox(0,0){$\Diamond$}}}
\put(1351,720){\raisebox{-.8pt}{\makebox(0,0){$\Diamond$}}}
\put(431,737){\raisebox{-.8pt}{\makebox(0,0){$\Diamond$}}}
\put(361,696){\makebox(0,0)[r]{$m=0.05$}}
\put(185,188){\makebox(0,0){$+$}}
\put(288,216){\makebox(0,0){$+$}}
\put(394,249){\makebox(0,0){$+$}}
\put(497,285){\makebox(0,0){$+$}}
\put(601,328){\makebox(0,0){$+$}}
\put(642,348){\makebox(0,0){$+$}}
\put(680,368){\makebox(0,0){$+$}}
\put(723,393){\makebox(0,0){$+$}}
\put(765,420){\makebox(0,0){$+$}}
\put(807,452){\makebox(0,0){$+$}}
\put(849,486){\makebox(0,0){$+$}}
\put(890,521){\makebox(0,0){$+$}}
\put(1016,596){\makebox(0,0){$+$}}
\put(1120,641){\makebox(0,0){$+$}}
\put(1230,680){\makebox(0,0){$+$}}
\put(1351,717){\makebox(0,0){$+$}}
\put(431,696){\makebox(0,0){$+$}}
\sbox{\plotpoint}{\rule[-0.400pt]{0.800pt}{0.800pt}}%
\put(361,655){\makebox(0,0)[r]{$m=0.00$}}
\put(185,142){\raisebox{-.8pt}{\makebox(0,0){$\Box$}}}
\put(288,172){\raisebox{-.8pt}{\makebox(0,0){$\Box$}}}
\put(394,208){\raisebox{-.8pt}{\makebox(0,0){$\Box$}}}
\put(497,248){\raisebox{-.8pt}{\makebox(0,0){$\Box$}}}
\put(601,295){\raisebox{-.8pt}{\makebox(0,0){$\Box$}}}
\put(642,317){\raisebox{-.8pt}{\makebox(0,0){$\Box$}}}
\put(680,339){\raisebox{-.8pt}{\makebox(0,0){$\Box$}}}
\put(723,367){\raisebox{-.8pt}{\makebox(0,0){$\Box$}}}
\put(765,398){\raisebox{-.8pt}{\makebox(0,0){$\Box$}}}
\put(807,434){\raisebox{-.8pt}{\makebox(0,0){$\Box$}}}
\put(849,474){\raisebox{-.8pt}{\makebox(0,0){$\Box$}}}
\put(890,515){\raisebox{-.8pt}{\makebox(0,0){$\Box$}}}
\put(1016,594){\raisebox{-.8pt}{\makebox(0,0){$\Box$}}}
\put(1120,639){\raisebox{-.8pt}{\makebox(0,0){$\Box$}}}
\put(1230,679){\raisebox{-.8pt}{\makebox(0,0){$\Box$}}}
\put(1351,715){\raisebox{-.8pt}{\makebox(0,0){$\Box$}}}
\put(431,655){\raisebox{-.8pt}{\makebox(0,0){$\Box$}}}
\end{picture}
\end{center}
\caption{$-S_f^{{\rm eff}}/V$ as a function of $E$ for $N_f=1$.}
\label{fig4}
\end{figure}
\begin{figure}[htb]
\begin{center}
\setlength{\unitlength}{0.240900pt}
\ifx\plotpoint\undefined\newsavebox{\plotpoint}\fi
\sbox{\plotpoint}{\rule[-0.200pt]{0.400pt}{0.400pt}}%
\begin{picture}(1500,900)(0,0)
\font\gnuplot=cmr10 at 10pt
\gnuplot
\sbox{\plotpoint}{\rule[-0.200pt]{0.400pt}{0.400pt}}%
\put(181.0,123.0){\rule[-0.200pt]{4.818pt}{0.400pt}}
\put(161,123){\makebox(0,0)[r]{-0.2}}
\put(1419.0,123.0){\rule[-0.200pt]{4.818pt}{0.400pt}}
\put(181.0,310.0){\rule[-0.200pt]{4.818pt}{0.400pt}}
\put(161,310){\makebox(0,0)[r]{0}}
\put(1419.0,310.0){\rule[-0.200pt]{4.818pt}{0.400pt}}
\put(181.0,497.0){\rule[-0.200pt]{4.818pt}{0.400pt}}
\put(161,497){\makebox(0,0)[r]{0.2}}
\put(1419.0,497.0){\rule[-0.200pt]{4.818pt}{0.400pt}}
\put(181.0,684.0){\rule[-0.200pt]{4.818pt}{0.400pt}}
\put(161,684){\makebox(0,0)[r]{0.4}}
\put(1419.0,684.0){\rule[-0.200pt]{4.818pt}{0.400pt}}
\put(181.0,123.0){\rule[-0.200pt]{0.400pt}{4.818pt}}
\put(181,82){\makebox(0,0){0}}
\put(181.0,757.0){\rule[-0.200pt]{0.400pt}{4.818pt}}
\put(433.0,123.0){\rule[-0.200pt]{0.400pt}{4.818pt}}
\put(433,82){\makebox(0,0){0.2}}
\put(433.0,757.0){\rule[-0.200pt]{0.400pt}{4.818pt}}
\put(684.0,123.0){\rule[-0.200pt]{0.400pt}{4.818pt}}
\put(684,82){\makebox(0,0){0.4}}
\put(684.0,757.0){\rule[-0.200pt]{0.400pt}{4.818pt}}
\put(936.0,123.0){\rule[-0.200pt]{0.400pt}{4.818pt}}
\put(936,82){\makebox(0,0){0.6}}
\put(936.0,757.0){\rule[-0.200pt]{0.400pt}{4.818pt}}
\put(1187.0,123.0){\rule[-0.200pt]{0.400pt}{4.818pt}}
\put(1187,82){\makebox(0,0){0.8}}
\put(1187.0,757.0){\rule[-0.200pt]{0.400pt}{4.818pt}}
\put(1439.0,123.0){\rule[-0.200pt]{0.400pt}{4.818pt}}
\put(1439,82){\makebox(0,0){1}}
\put(1439.0,757.0){\rule[-0.200pt]{0.400pt}{4.818pt}}
\put(181.0,123.0){\rule[-0.200pt]{303.052pt}{0.400pt}}
\put(1439.0,123.0){\rule[-0.200pt]{0.400pt}{157.549pt}}
\put(181.0,777.0){\rule[-0.200pt]{303.052pt}{0.400pt}}
\put(40,450){\makebox(0,0){$-{S^{\rm{eff}}_f \over V}$}}
\put(810,21){\makebox(0,0){$E$}}
\put(810,839){\makebox(0,0){$N_f=2$}}
\put(181.0,123.0){\rule[-0.200pt]{0.400pt}{157.549pt}}
\put(361,737){\makebox(0,0)[r]{$m=0.10$}}
\put(185,233){\raisebox{-.8pt}{\makebox(0,0){$\Diamond$}}}
\put(288,259){\raisebox{-.8pt}{\makebox(0,0){$\Diamond$}}}
\put(394,290){\raisebox{-.8pt}{\makebox(0,0){$\Diamond$}}}
\put(497,323){\raisebox{-.8pt}{\makebox(0,0){$\Diamond$}}}
\put(601,362){\raisebox{-.8pt}{\makebox(0,0){$\Diamond$}}}
\put(642,380){\raisebox{-.8pt}{\makebox(0,0){$\Diamond$}}}
\put(680,398){\raisebox{-.8pt}{\makebox(0,0){$\Diamond$}}}
\put(723,420){\raisebox{-.8pt}{\makebox(0,0){$\Diamond$}}}
\put(765,445){\raisebox{-.8pt}{\makebox(0,0){$\Diamond$}}}
\put(807,472){\raisebox{-.8pt}{\makebox(0,0){$\Diamond$}}}
\put(849,502){\raisebox{-.8pt}{\makebox(0,0){$\Diamond$}}}
\put(890,533){\raisebox{-.8pt}{\makebox(0,0){$\Diamond$}}}
\put(1016,602){\raisebox{-.8pt}{\makebox(0,0){$\Diamond$}}}
\put(1120,646){\raisebox{-.8pt}{\makebox(0,0){$\Diamond$}}}
\put(1230,684){\raisebox{-.8pt}{\makebox(0,0){$\Diamond$}}}
\put(1351,720){\raisebox{-.8pt}{\makebox(0,0){$\Diamond$}}}
\put(431,737){\raisebox{-.8pt}{\makebox(0,0){$\Diamond$}}}
\put(361,696){\makebox(0,0)[r]{$m=0.05$}}
\put(185,188){\makebox(0,0){$+$}}
\put(288,216){\makebox(0,0){$+$}}
\put(394,249){\makebox(0,0){$+$}}
\put(497,285){\makebox(0,0){$+$}}
\put(601,329){\makebox(0,0){$+$}}
\put(642,348){\makebox(0,0){$+$}}
\put(680,368){\makebox(0,0){$+$}}
\put(723,393){\makebox(0,0){$+$}}
\put(765,421){\makebox(0,0){$+$}}
\put(807,452){\makebox(0,0){$+$}}
\put(849,486){\makebox(0,0){$+$}}
\put(890,521){\makebox(0,0){$+$}}
\put(1016,596){\makebox(0,0){$+$}}
\put(1120,641){\makebox(0,0){$+$}}
\put(1230,680){\makebox(0,0){$+$}}
\put(1351,717){\makebox(0,0){$+$}}
\put(431,696){\makebox(0,0){$+$}}
\sbox{\plotpoint}{\rule[-0.400pt]{0.800pt}{0.800pt}}%
\put(361,655){\makebox(0,0)[r]{$m=0.00$}}
\put(185,142){\raisebox{-.8pt}{\makebox(0,0){$\Box$}}}
\put(288,173){\raisebox{-.8pt}{\makebox(0,0){$\Box$}}}
\put(394,208){\raisebox{-.8pt}{\makebox(0,0){$\Box$}}}
\put(497,248){\raisebox{-.8pt}{\makebox(0,0){$\Box$}}}
\put(601,295){\raisebox{-.8pt}{\makebox(0,0){$\Box$}}}
\put(642,317){\raisebox{-.8pt}{\makebox(0,0){$\Box$}}}
\put(680,340){\raisebox{-.8pt}{\makebox(0,0){$\Box$}}}
\put(723,367){\raisebox{-.8pt}{\makebox(0,0){$\Box$}}}
\put(765,398){\raisebox{-.8pt}{\makebox(0,0){$\Box$}}}
\put(807,434){\raisebox{-.8pt}{\makebox(0,0){$\Box$}}}
\put(849,474){\raisebox{-.8pt}{\makebox(0,0){$\Box$}}}
\put(890,515){\raisebox{-.8pt}{\makebox(0,0){$\Box$}}}
\put(1016,594){\raisebox{-.8pt}{\makebox(0,0){$\Box$}}}
\put(1120,639){\raisebox{-.8pt}{\makebox(0,0){$\Box$}}}
\put(1230,679){\raisebox{-.8pt}{\makebox(0,0){$\Box$}}}
\put(1351,715){\raisebox{-.8pt}{\makebox(0,0){$\Box$}}}
\put(431,655){\raisebox{-.8pt}{\makebox(0,0){$\Box$}}}
\end{picture}
\end{center}
\caption{$-S_f^{{\rm eff}}/V$ as a function of $E$  for $N_f=2$.}
\label{fig5}
\end{figure}
\begin{figure}[htb]
\begin{center}
\setlength{\unitlength}{0.240900pt}
\ifx\plotpoint\undefined\newsavebox{\plotpoint}\fi
\sbox{\plotpoint}{\rule[-0.200pt]{0.400pt}{0.400pt}}%
\begin{picture}(1500,900)(0,0)
\font\gnuplot=cmr10 at 10pt
\gnuplot
\sbox{\plotpoint}{\rule[-0.200pt]{0.400pt}{0.400pt}}%
\put(181.0,123.0){\rule[-0.200pt]{4.818pt}{0.400pt}}
\put(161,123){\makebox(0,0)[r]{-0.3}}
\put(1419.0,123.0){\rule[-0.200pt]{4.818pt}{0.400pt}}
\put(181.0,310.0){\rule[-0.200pt]{4.818pt}{0.400pt}}
\put(161,310){\makebox(0,0)[r]{0}}
\put(1419.0,310.0){\rule[-0.200pt]{4.818pt}{0.400pt}}
\put(181.0,497.0){\rule[-0.200pt]{4.818pt}{0.400pt}}
\put(161,497){\makebox(0,0)[r]{0.3}}
\put(1419.0,497.0){\rule[-0.200pt]{4.818pt}{0.400pt}}
\put(181.0,684.0){\rule[-0.200pt]{4.818pt}{0.400pt}}
\put(161,684){\makebox(0,0)[r]{0.6}}
\put(1419.0,684.0){\rule[-0.200pt]{4.818pt}{0.400pt}}
\put(181.0,123.0){\rule[-0.200pt]{0.400pt}{4.818pt}}
\put(181,82){\makebox(0,0){0}}
\put(181.0,757.0){\rule[-0.200pt]{0.400pt}{4.818pt}}
\put(433.0,123.0){\rule[-0.200pt]{0.400pt}{4.818pt}}
\put(433,82){\makebox(0,0){0.2}}
\put(433.0,757.0){\rule[-0.200pt]{0.400pt}{4.818pt}}
\put(684.0,123.0){\rule[-0.200pt]{0.400pt}{4.818pt}}
\put(684,82){\makebox(0,0){0.4}}
\put(684.0,757.0){\rule[-0.200pt]{0.400pt}{4.818pt}}
\put(936.0,123.0){\rule[-0.200pt]{0.400pt}{4.818pt}}
\put(936,82){\makebox(0,0){0.6}}
\put(936.0,757.0){\rule[-0.200pt]{0.400pt}{4.818pt}}
\put(1187.0,123.0){\rule[-0.200pt]{0.400pt}{4.818pt}}
\put(1187,82){\makebox(0,0){0.8}}
\put(1187.0,757.0){\rule[-0.200pt]{0.400pt}{4.818pt}}
\put(1439.0,123.0){\rule[-0.200pt]{0.400pt}{4.818pt}}
\put(1439,82){\makebox(0,0){1}}
\put(1439.0,757.0){\rule[-0.200pt]{0.400pt}{4.818pt}}
\put(181.0,123.0){\rule[-0.200pt]{303.052pt}{0.400pt}}
\put(1439.0,123.0){\rule[-0.200pt]{0.400pt}{157.549pt}}
\put(181.0,777.0){\rule[-0.200pt]{303.052pt}{0.400pt}}
\put(40,450){\makebox(0,0){$-{S^{\rm{eff}}_f\over V}$}}
\put(810,21){\makebox(0,0){$E$}}
\put(810,839){\makebox(0,0){$N_f=3$}}
\put(181.0,123.0){\rule[-0.200pt]{0.400pt}{157.549pt}}
\put(361,737){\makebox(0,0)[r]{$m=0.10$}}
\put(185,233){\raisebox{-.8pt}{\makebox(0,0){$\Diamond$}}}
\put(288,259){\raisebox{-.8pt}{\makebox(0,0){$\Diamond$}}}
\put(394,290){\raisebox{-.8pt}{\makebox(0,0){$\Diamond$}}}
\put(497,323){\raisebox{-.8pt}{\makebox(0,0){$\Diamond$}}}
\put(601,362){\raisebox{-.8pt}{\makebox(0,0){$\Diamond$}}}
\put(642,380){\raisebox{-.8pt}{\makebox(0,0){$\Diamond$}}}
\put(680,399){\raisebox{-.8pt}{\makebox(0,0){$\Diamond$}}}
\put(723,420){\raisebox{-.8pt}{\makebox(0,0){$\Diamond$}}}
\put(765,445){\raisebox{-.8pt}{\makebox(0,0){$\Diamond$}}}
\put(807,473){\raisebox{-.8pt}{\makebox(0,0){$\Diamond$}}}
\put(849,502){\raisebox{-.8pt}{\makebox(0,0){$\Diamond$}}}
\put(890,533){\raisebox{-.8pt}{\makebox(0,0){$\Diamond$}}}
\put(1016,603){\raisebox{-.8pt}{\makebox(0,0){$\Diamond$}}}
\put(1120,646){\raisebox{-.8pt}{\makebox(0,0){$\Diamond$}}}
\put(1230,684){\raisebox{-.8pt}{\makebox(0,0){$\Diamond$}}}
\put(1351,720){\raisebox{-.8pt}{\makebox(0,0){$\Diamond$}}}
\put(431,737){\raisebox{-.8pt}{\makebox(0,0){$\Diamond$}}}
\put(361,696){\makebox(0,0)[r]{$m=0.05$}}
\put(185,188){\makebox(0,0){$+$}}
\put(288,216){\makebox(0,0){$+$}}
\put(394,249){\makebox(0,0){$+$}}
\put(497,285){\makebox(0,0){$+$}}
\put(601,329){\makebox(0,0){$+$}}
\put(642,348){\makebox(0,0){$+$}}
\put(680,369){\makebox(0,0){$+$}}
\put(723,393){\makebox(0,0){$+$}}
\put(765,421){\makebox(0,0){$+$}}
\put(807,452){\makebox(0,0){$+$}}
\put(849,486){\makebox(0,0){$+$}}
\put(890,521){\makebox(0,0){$+$}}
\put(1016,596){\makebox(0,0){$+$}}
\put(1120,641){\makebox(0,0){$+$}}
\put(1230,680){\makebox(0,0){$+$}}
\put(1351,717){\makebox(0,0){$+$}}
\put(431,696){\makebox(0,0){$+$}}
\sbox{\plotpoint}{\rule[-0.400pt]{0.800pt}{0.800pt}}%
\put(361,655){\makebox(0,0)[r]{$m=0.00$}}
\put(185,142){\raisebox{-.8pt}{\makebox(0,0){$\Box$}}}
\put(288,173){\raisebox{-.8pt}{\makebox(0,0){$\Box$}}}
\put(394,209){\raisebox{-.8pt}{\makebox(0,0){$\Box$}}}
\put(497,248){\raisebox{-.8pt}{\makebox(0,0){$\Box$}}}
\put(601,296){\raisebox{-.8pt}{\makebox(0,0){$\Box$}}}
\put(642,317){\raisebox{-.8pt}{\makebox(0,0){$\Box$}}}
\put(680,340){\raisebox{-.8pt}{\makebox(0,0){$\Box$}}}
\put(723,367){\raisebox{-.8pt}{\makebox(0,0){$\Box$}}}
\put(765,399){\raisebox{-.8pt}{\makebox(0,0){$\Box$}}}
\put(807,435){\raisebox{-.8pt}{\makebox(0,0){$\Box$}}}
\put(849,475){\raisebox{-.8pt}{\makebox(0,0){$\Box$}}}
\put(890,515){\raisebox{-.8pt}{\makebox(0,0){$\Box$}}}
\put(1016,594){\raisebox{-.8pt}{\makebox(0,0){$\Box$}}}
\put(1120,639){\raisebox{-.8pt}{\makebox(0,0){$\Box$}}}
\put(1230,679){\raisebox{-.8pt}{\makebox(0,0){$\Box$}}}
\put(1351,716){\raisebox{-.8pt}{\makebox(0,0){$\Box$}}}
\put(431,655){\raisebox{-.8pt}{\makebox(0,0){$\Box$}}}
\end{picture}
\end{center}
\caption{$-S_f^{{\rm eff}}/V$ as a function of $E$  for $N_f=3$.}
\label{fig6}
\end{figure}
\begin{figure}[htb]
\begin{center}
\setlength{\unitlength}{0.240900pt}
\ifx\plotpoint\undefined\newsavebox{\plotpoint}\fi
\sbox{\plotpoint}{\rule[-0.200pt]{0.400pt}{0.400pt}}%
\begin{picture}(1500,900)(0,0)
\font\gnuplot=cmr10 at 10pt
\gnuplot
\sbox{\plotpoint}{\rule[-0.200pt]{0.400pt}{0.400pt}}%
\put(181.0,123.0){\rule[-0.200pt]{4.818pt}{0.400pt}}
\put(161,123){\makebox(0,0)[r]{-0.4}}
\put(1419.0,123.0){\rule[-0.200pt]{4.818pt}{0.400pt}}
\put(181.0,310.0){\rule[-0.200pt]{4.818pt}{0.400pt}}
\put(161,310){\makebox(0,0)[r]{0}}
\put(1419.0,310.0){\rule[-0.200pt]{4.818pt}{0.400pt}}
\put(181.0,497.0){\rule[-0.200pt]{4.818pt}{0.400pt}}
\put(161,497){\makebox(0,0)[r]{0.4}}
\put(1419.0,497.0){\rule[-0.200pt]{4.818pt}{0.400pt}}
\put(181.0,684.0){\rule[-0.200pt]{4.818pt}{0.400pt}}
\put(161,684){\makebox(0,0)[r]{0.8}}
\put(1419.0,684.0){\rule[-0.200pt]{4.818pt}{0.400pt}}
\put(181.0,123.0){\rule[-0.200pt]{0.400pt}{4.818pt}}
\put(181,82){\makebox(0,0){0}}
\put(181.0,757.0){\rule[-0.200pt]{0.400pt}{4.818pt}}
\put(433.0,123.0){\rule[-0.200pt]{0.400pt}{4.818pt}}
\put(433,82){\makebox(0,0){0.2}}
\put(433.0,757.0){\rule[-0.200pt]{0.400pt}{4.818pt}}
\put(684.0,123.0){\rule[-0.200pt]{0.400pt}{4.818pt}}
\put(684,82){\makebox(0,0){0.4}}
\put(684.0,757.0){\rule[-0.200pt]{0.400pt}{4.818pt}}
\put(936.0,123.0){\rule[-0.200pt]{0.400pt}{4.818pt}}
\put(936,82){\makebox(0,0){0.6}}
\put(936.0,757.0){\rule[-0.200pt]{0.400pt}{4.818pt}}
\put(1187.0,123.0){\rule[-0.200pt]{0.400pt}{4.818pt}}
\put(1187,82){\makebox(0,0){0.8}}
\put(1187.0,757.0){\rule[-0.200pt]{0.400pt}{4.818pt}}
\put(1439.0,123.0){\rule[-0.200pt]{0.400pt}{4.818pt}}
\put(1439,82){\makebox(0,0){1}}
\put(1439.0,757.0){\rule[-0.200pt]{0.400pt}{4.818pt}}
\put(181.0,123.0){\rule[-0.200pt]{303.052pt}{0.400pt}}
\put(1439.0,123.0){\rule[-0.200pt]{0.400pt}{157.549pt}}
\put(181.0,777.0){\rule[-0.200pt]{303.052pt}{0.400pt}}
\put(40,450){\makebox(0,0){$-{S^{\rm{eff}}_f \over V}$}}
\put(810,21){\makebox(0,0){$E$}}
\put(810,839){\makebox(0,0){$N_f=4$}}
\put(181.0,123.0){\rule[-0.200pt]{0.400pt}{157.549pt}}
\put(361,737){\makebox(0,0)[r]{$m=0.10$}}
\put(185,233){\raisebox{-.8pt}{\makebox(0,0){$\Diamond$}}}
\put(288,259){\raisebox{-.8pt}{\makebox(0,0){$\Diamond$}}}
\put(394,290){\raisebox{-.8pt}{\makebox(0,0){$\Diamond$}}}
\put(497,323){\raisebox{-.8pt}{\makebox(0,0){$\Diamond$}}}
\put(601,362){\raisebox{-.8pt}{\makebox(0,0){$\Diamond$}}}
\put(642,380){\raisebox{-.8pt}{\makebox(0,0){$\Diamond$}}}
\put(680,399){\raisebox{-.8pt}{\makebox(0,0){$\Diamond$}}}
\put(723,421){\raisebox{-.8pt}{\makebox(0,0){$\Diamond$}}}
\put(765,445){\raisebox{-.8pt}{\makebox(0,0){$\Diamond$}}}
\put(807,473){\raisebox{-.8pt}{\makebox(0,0){$\Diamond$}}}
\put(849,502){\raisebox{-.8pt}{\makebox(0,0){$\Diamond$}}}
\put(890,533){\raisebox{-.8pt}{\makebox(0,0){$\Diamond$}}}
\put(1016,603){\raisebox{-.8pt}{\makebox(0,0){$\Diamond$}}}
\put(1120,646){\raisebox{-.8pt}{\makebox(0,0){$\Diamond$}}}
\put(1230,684){\raisebox{-.8pt}{\makebox(0,0){$\Diamond$}}}
\put(1351,720){\raisebox{-.8pt}{\makebox(0,0){$\Diamond$}}}
\put(431,737){\raisebox{-.8pt}{\makebox(0,0){$\Diamond$}}}
\put(361,696){\makebox(0,0)[r]{$m=0.05$}}
\put(185,188){\makebox(0,0){$+$}}
\put(288,216){\makebox(0,0){$+$}}
\put(394,249){\makebox(0,0){$+$}}
\put(497,285){\makebox(0,0){$+$}}
\put(601,329){\makebox(0,0){$+$}}
\put(642,348){\makebox(0,0){$+$}}
\put(680,369){\makebox(0,0){$+$}}
\put(723,393){\makebox(0,0){$+$}}
\put(765,421){\makebox(0,0){$+$}}
\put(807,452){\makebox(0,0){$+$}}
\put(849,487){\makebox(0,0){$+$}}
\put(890,521){\makebox(0,0){$+$}}
\put(1016,597){\makebox(0,0){$+$}}
\put(1120,641){\makebox(0,0){$+$}}
\put(1230,680){\makebox(0,0){$+$}}
\put(1351,717){\makebox(0,0){$+$}}
\put(431,696){\makebox(0,0){$+$}}
\sbox{\plotpoint}{\rule[-0.400pt]{0.800pt}{0.800pt}}%
\put(361,655){\makebox(0,0)[r]{$m=0.00$}}
\put(185,142){\raisebox{-.8pt}{\makebox(0,0){$\Box$}}}
\put(288,173){\raisebox{-.8pt}{\makebox(0,0){$\Box$}}}
\put(394,209){\raisebox{-.8pt}{\makebox(0,0){$\Box$}}}
\put(497,248){\raisebox{-.8pt}{\makebox(0,0){$\Box$}}}
\put(601,296){\raisebox{-.8pt}{\makebox(0,0){$\Box$}}}
\put(642,317){\raisebox{-.8pt}{\makebox(0,0){$\Box$}}}
\put(680,340){\raisebox{-.8pt}{\makebox(0,0){$\Box$}}}
\put(723,367){\raisebox{-.8pt}{\makebox(0,0){$\Box$}}}
\put(765,399){\raisebox{-.8pt}{\makebox(0,0){$\Box$}}}
\put(807,435){\raisebox{-.8pt}{\makebox(0,0){$\Box$}}}
\put(849,475){\raisebox{-.8pt}{\makebox(0,0){$\Box$}}}
\put(890,515){\raisebox{-.8pt}{\makebox(0,0){$\Box$}}}
\put(1016,594){\raisebox{-.8pt}{\makebox(0,0){$\Box$}}}
\put(1120,639){\raisebox{-.8pt}{\makebox(0,0){$\Box$}}}
\put(1230,679){\raisebox{-.8pt}{\makebox(0,0){$\Box$}}}
\put(1351,716){\raisebox{-.8pt}{\makebox(0,0){$\Box$}}}
\put(431,655){\raisebox{-.8pt}{\makebox(0,0){$\Box$}}}
\end{picture}
\end{center}
\caption{$-S_f^{{\rm eff}}/V$ as a function of $E$  for $N_f=4$.}
\label{fig7}
\end{figure}
\input fig8.tex
\input fig9.tex
\input fig10.tex
\input fig11.tex

\end{document}